# Absolute ion detection efficiencies of microchannel plates and funnel microchannel plates for multi-coincidence detection

K. Fehre[*,1], D. Trojanowskaja[*], J. Gatzke[*], M. Kunitski[*], F. Trinter[*], S. Zeller[*], L. Ph. H. Schmidt[*], J. Stohner[†], R. Berger[‡], A. Czasch[§], O. Jagutzki[§], T. Jahnke[*], R. Dörner[*,1] and M. S. Schöffler[*,1]

[*]Institut für Kernphysik Goethe-Universität Frankfurt, Max-von-Laue-Str. 1, DE-60438 Germany
[†] Institute of Chemistry and Biotechnology, Zurich University of Applied Sciences, Campus Riedbach, Einsiedlerstr. 31, CH-8820 Wädenswil, Switzerland
[§]RoentDek Handels GmbH, Im Vogelshaag 8, DE-56779 Kelkheim, Germany
[‡]Philipps-Universität Marburg, Hans-Meerwein-Straße 4, DE-35032 Marburg, Germany

Modern momentum imaging techniques allow for the investigation of complex molecules in the gas phase by detection of several fragment ions in coincidence. For these studies, it is of great importance that the single-particle detection efficiency ε is as high as possible, as the overall efficiency scales with $\varepsilon^n$, i.e. the power of the number of detected particles. Here we present measured absolute detection efficiencies for protons of several micro-channel plates (MCPs), including efficiency enhanced "funnel MCPs". Furthermore, the relative detection efficiency for two-, three-, four-, and five-body fragmentation of CHBrClF has been examined. The "funnel" MCPs exhibit an efficiency of approx. 90 %, gaining a factor of 24 (as compared to "normal" MCPs) in case of a five-fold ion coincidence detection.

**Introduction**

Micro-channel plate (MCP) electron multipliers are much-used detectors for photons and charged particles of low energy today. One major application is their use for single-particle detection where the position of impact and the arrival time is obtained [1]. For many experiments, the absolute detection efficiency ε of an MCP is an important parameter. It describes the probability that an impacting particle (photon, electron or ion) triggers a signal on the detector. In case of coincident multiple particle detection a high efficiency is particularly essential, as the efficiency for detecting n particles scales with $\varepsilon^n$. Accordingly, experimental approaches that typically benefit from increased MCP detection efficiencies range from the magnetic bottle time-of-flight technique [2] to the Cold Target Recoil Ion Momentum Spectroscopy (COLTRIMS) [3]. The efficiency of an MCP depends on various factors such as the mass and the kinetic energy of the particles to be detected. Studies by Krems et al. [4] show that the upper limit of the efficiency is given by the open area ratio (OAR) of the channel plate, as particles are hardly recognized when they do not impinge into a pore of the MCP [5]. While larger pores and/or thinner walls increase the OAR, they result in a reduced timing resolution [6] and cause increased ion feedback. However, as a minimum wall thickness of the pores has to be maintained to ensure mechanical stability, novel approaches to expand the pores only near the surface have been developed: By means of new etching methods so called "funnel" MCPs [7] with greatly increased OAR of up to 90 % are nowadays commercially available and several experiments have already demonstrated their increased efficiency [8].

Typically, the MCP efficiency is determined by comparison to other detection schemes [9–14], as for example, to measured ion currents in a Faraday cup. So far, only few experiments determined the absolute detection efficiency directly [15, 16]. In this article, a method similar to one in [15] has been employed in order to achieve this: By measuring the ratio of two contributions which occur intrinsically in the reaction examined by the MCP detector, its efficiency can be deduced without need for an external reference measurement.

To this end the double electron capture from $H_2$ into a fast doubly charged argon projectile (20 keV/u $Ar^{2+}$ + $H_2$ → $Ar^0$ + $H^+$ + $H^+$) was utilized to create a pair of protons. By triggering on neutralized $Ar^0$ projectiles in our experiment, we are able to identify those events where two protons were created.

Those events can now be divided into cases in which the proton arriving first at the MCP is detected and cases in which the first one is missed, but the second proton is detected. From the ratio of these data subsets the proton detection efficiency can be directly deduced. Using this scheme, we measured the absolute detection efficiency of three different types of MCPs; details are given in Table 1. Furthermore, in order to demonstrate the importance of a high detection efficiency in multi-coinci-

---

dent measurements, we compare an efficiency-enhanced "funnel" MCP to a "standard" MCP under identical conditions in an experiment in which we multiply ionize CHBrClF employing a strong femtosecond-laser resulting in up to five ionic fragments to be detected.

**Determining the absolute MCP detection efficiency for protons**

In order to determine the absolute MCP detection efficiency for a proton, Coulomb explosion of $H_2$ molecules has been examined using the COLTRIMS technology [3]. After double ionization, the molecule breaks up into two protons which are emitted back-to-back. The double ionization has been triggered utilizing a beam of doubly charged argon ions: 20 keV/u $Ar^{2+}$ + $H_2$ → $Ar^0$ + $H^+$ + $H^+$. $Ar^+$ ions are accelerated by a Van de Graaff accelerator (800 keV, corresponding to 20 keV/u) and stripped down to $Ar^{2+}$ by passing through a gas cell. This projectile beam is then intersected with a molecular jet of $H_2$ (target density = $5·10^{10}$ molecules/cm²) at right angle. We clean the $Ar^{2+}$ beam from impurities (i.e. $Ar^+$ and $Ar^0$) using a vertical electrostatic deflector (Fig. 1), only projectiles entering the reaction chamber as $Ar^{2+}$ reach the projectile detector. This scheme encodes the Ar charge state before entering the reaction chamber in the y-axis.

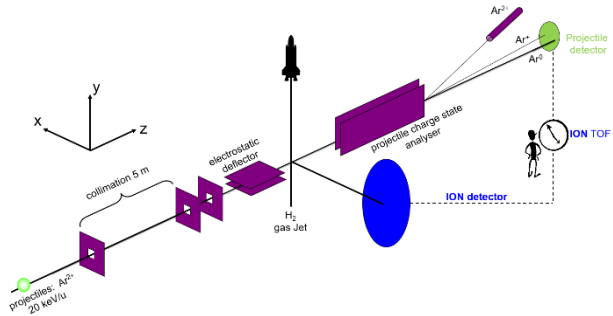

**Fig. 1** Schematic of the COLTRIMS experiment. The $Ar^{2+}$ ion beam from a Van de Graaff accelerator passes ion optical elements and enters the reaction chamber where it is crossed with a $H_2$ molecular beam. The arrival times and positions of impact of the reaction products are measured by two MCP detectors.

When the argon ion captures two electrons from the $H_2$ molecule, two $H^+$ ions must have been created. The neutralized $Ar^0$ atom is separated from the main beam by a horizontally mounted projectile charge state analyzer. These charge-exchanged projectiles are detected by a position- and time-sensitive micro channel plate detector. The protons are driven by a weak electric field (E = 100 V/cm) over 20 cm, finally hitting a second microchannel plate detector, equipped with a hexagonal delay-line anode for position read-out. Every detector consists out of two MCP plates: For further amplification, a second MCP (Photonis) was mounted behind the first MCP.

The main chamber was baked, resulting in a residual gas pressure (without gas jet) of $2·10^{-7}$ Pa.

The transfer of the two electrons from the $H_2$ molecule from the bound state of the projectile takes place on a very short time scale; the nuclei are quasi frozen. The two protons are now being driven in opposite directions due to the Coulomb repulsion. The kinetic energy release (KER) gained in the Coulomb explosion is known to be approx. 19 eV as it corresponds to the internuclear distance of the two hydrogen atoms in the ground state of the molecule. The proton which is emitted towards the ion detector is detected first, while the other proton (initially heading away from the detector) travels longer. It is at first decelerated by the spectrometer's electric field and then driven back towards the ion detector, as well. The protons' times-of-flight corresponds in good approximation to the spatial orientation of the molecule at the instant of double ionization: Shortest and longest flight times belong to cases where the $H_2$ molecule is oriented in parallel with respect to the electric field of the spectrometer, while equal times-of-flight of both protons occurs in events where the molecule was oriented perpendicular to electric field direction. Accordingly, the first detected proton is expected to have a time-of-flight (TOF) distribution from 600 ns to 650 ns (for the given spectrometer geometry and electric field), while the proton which is detected second is supposed to occur at 650 ns < TOF < 700 ns. The measured TOF distribution of the first hit on the detector is shown in Figures 2c,d). Other than expected the measured TOF distribution spans an overall range from 600 ns to 700 ns. It turns out, that this is due to imperfect detection efficiency: if the first proton (that started towards the detector) has been missed then the second proton (initially heading away from the detector) is detected instead as the "first" ion. Correspondingly, the absolute detection efficiency can deduced from the TOF distribution $g_1(t)$ of the first measured ion applying the following equation:

$$\varepsilon = 1 - \frac{\int_{650\,ns}^{700\,ns} g_1(t)\,dt}{\int_{600\,ns}^{650\,ns} g_1(t)\,dt}$$

In order to derive the MCP efficiency more accurately, only events with a measured 14 eV < KER < 25 eV (corresponding to a range for the linear momentum $p$ of 30.7 < |p/a.u.| < 41) were taken into account. This eliminates, for example, false coincidences where just one electron was removed from the $H_2$ molecule causing it to rapidly dissociate into $H^+$ and $H^0$. The latter events yield very low proton kinetic energies. Fig. 2, top depicts the measured proton linear momenta. Additionally, a separate measurement of the background was made as the $Ar^{2+}$ beam can interact with the residual gas in the chamber anywhere along the beam path. Therefore, the projectile beam was moved a few millimeters aside to avoid crossing of the gas jet. In the calculation of linear momenta of the detected particles, both the position of the reaction volume and the time at which projectile and target collide in this reaction volume enter the equations. Products formed in reactions occurring along the projectile beam but outside this predefined crossing region of target beam and gas jet are assigned with false linear momenta, by which background reactions bypass the KER gate.

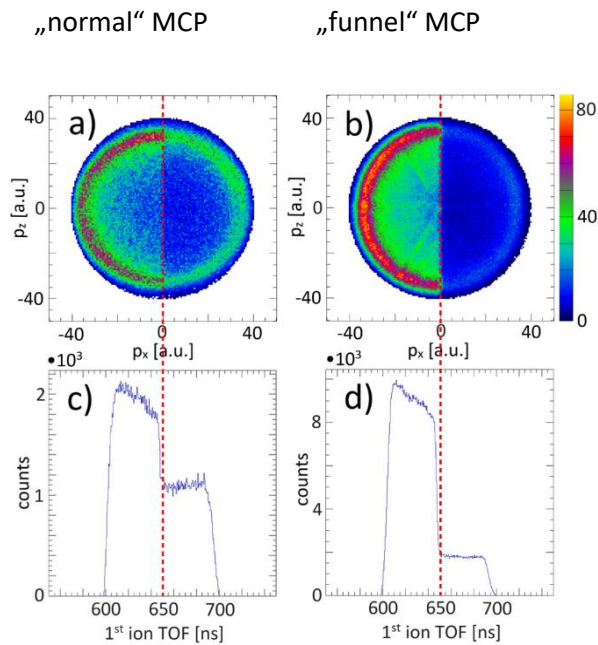

**Fig. 2** Upper row: Measured proton momenta for 30.7 a.u < |p| < 41 a.u in direction of the projectile beam ($p_z$) and the spectrometer's electric field direction ($p_x$). Lower row: Time-of-flight of the first detected ion. Left: A large share of missed first $H^+$ ions indicate a comparably poor detection efficiency (measured 56%). Right: Same as left, but for efficiency-enhanced MCP (measured 86%) – (background subtracted in both cases).

The background was dominated in case of the 60 % OAR MCP (see Table 1 for details) by water and for the other two MCPs (70 % and 90 % OAR) by residual $H_2$. The latter contribution is trickier to evaluate as the deduction of the proton momenta from the measured flight times and impact position on the detector relies on a small reaction volume located at a known position. For protons generated elsewhere it will result in wrong values. Accordingly, the background contribution produces a much wider $H_2$-KER distribution. By investigating this distribution in the background measurement we were able to subtract it from the main datasets; the influence of this correction and the resulting overall efficiency of the MCPs under test can be seen in Table 1.

In Fig. 2 a) the proton's linear momentum is shown in the TOF direction ($p_x$) and in the direction parallel to the $Ar^{2+}$ beam ($p_z$) after gating on $Ar^0$ atoms. As only the first hit is displayed, a detector with 100 % efficiency would show solely the left half of the sphere ($p_x<0$). The ratio between the signals for $p_x>0$ and $p_x<0$ directly leads to the absolute detection efficiency. Fig. 2 c) displays the TOF corresponding to the first detected hit directly. As indicated by the dotted red line, this representation can be understood as a projection from the upper panels. While figure Fig. 2 b) and d) were measured with the efficiency enhanced funnel MCP, Fig. 2 a) and c) were measured for the "standard" MCP (60 % OAR). The larger share of missed first $H^+$ ions (figure 2 c) and therefore the larger contribution of TOFs>650 ns, indicates a comparably poor detection efficiency compared to Fig. 2 d) (see Table 1 for numbers).

It is known, that the quantum efficiency depends strongly on the impact energy of the particles **[4]**. To ensure proper saturation of the quantum efficiency of the MCPs, the absolute efficiency was measured additionally for one MCP type as a function of the proton's impact energy (Fig. 3). The impact energy on the detector was altered by changing the electric field of the COLTRIMS spectrometer. Only a small change in efficiency over a wide range of the acceleration voltage is visible, suggesting that the applied voltage was sufficient to reach saturation. Further examination pointed out that the slight increase in efficiency towards higher proton kinetic energies is most probably not connected to the MCP's quantum efficiency but to a change in the MCP's pulse height distribution. If the overall MCP pulse height increases, less pulses are discarded due to the threshold set to discriminate the real signal against electronic noise. To take also this small effect into account for the determination of the absolute detector efficiency, the pulse height distribution (shown in Fig. 4) was recorded in all experiments and was considered in the calculation of the efficiencies. We have identified the red area (which represents pulses that were discriminated by the electronics) to be a few percent (detailed numbers of each MCP, are shown Table 1, last column).

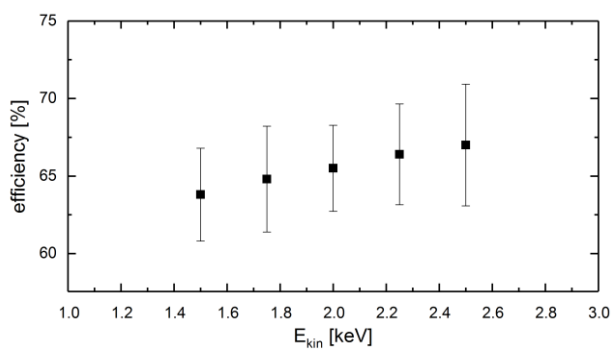

**Fig. 3** $E_{kin}$ [keV] versus the quantum efficiency for the Hamamatsu MCP with an OAR of 70 % (only background subtracted for 2 keV, no pulse height correction).

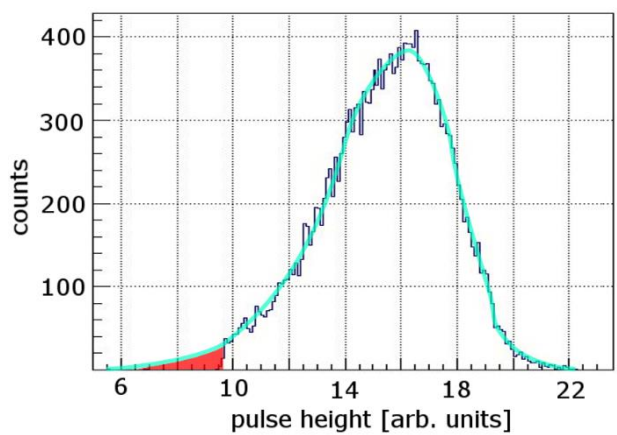

**Fig. 4** Integral pulse height distribution of the "funnel" MCP. Measured data are plotted in dark blue; the semi-transparent turquoise line was added to guide the eye. Only pulses higher than 9.6 arb. units can be recorded as smaller pulses will not stand out from the noise. Events which are not accessible by the experiment are marked by the red area.

A further effect we studied is the dependence of the detection efficiency on the angle of incidence on the MCP. Since the pores in the MCP are at a small bias angle to the surface normal (8°- 20°) and the particles pass from the reaction zone on parabolic trajectories to the MCP, the particles hit different locations on the MCP at a different angle relative to the MCP pores (see Fig. 5, top for a sketch). We observe that this effect leads to a change in the detection efficiency of about 3% (for the "funnel" MCP), see Fig. 5, bottom. The deviation of the efficiency from the red line for θ close to 0° (see Fig. 5) is not due to experimental errors, but only shows that the dependence of the efficiency on φ has a more complex relationship.

The absolute detection efficiency as displayed in Table 1 therefore indicates the value averaged over this effect.

One can expect that this impact angle dependence of the efficiency is also influenced by the shape of the pore.

While the "funnel" MCP shows a relative deviation in efficiency $(\varepsilon_{max}-\varepsilon_{min})/\varepsilon$ of 0.02, the 70 % OAR MCP shows for the same inclination of the pores a relative deviation of 0.05, derived from comparison of spectra similar to Fig. 5 (not shown here).

Table 1 summarizes the measured absolute detection efficiencies for all MCPs under investigation and shows how larger open area ratios enhance the detection efficiency.

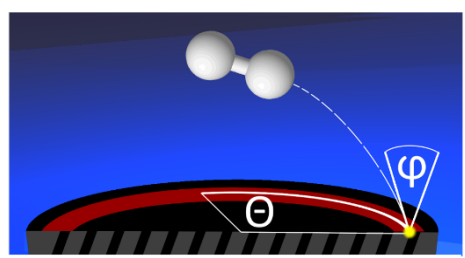

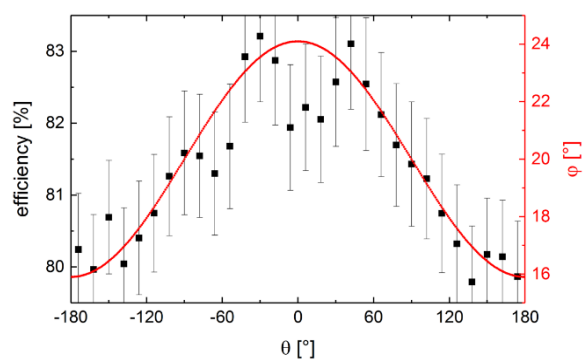

**Fig. 5** Dependence of the ion detection efficiency on the angle of incidence. The drawing in the upper panel illustrates why the ion hits the detector at different angles relative to the pores of the MCP and explains the applied angles. The KER gate is marked in red for molecular orientations in which the molecular axis is almost parallel to the MCP surface. Depending on the impact on the detector, a particle that starts in the reaction zone has a different relative angle to the pores in the MCP. The detector efficiency was determined for different sections of θ. The red y-axis to the right in the lower panel indicates a relative angle of incidence φ revealing the direct relationship between the angle of incidence and the detector efficiency.

|  | OAR [%] | Measured apparent efficiency before background correction [%] | Measured efficiency [%] (background corrected) | Measured efficiency [%] (background corrected with pulse height distribution) |
|---|---|---|---|---|
| Hamamatsu ("funnel") 12 µm; 60:1; 20° | 90 | 79.6 ± 0.4 | 83.4 ± 1.0 | 86.0 ± 1.2 |
| Hamamatsu 12 µm; 80:1; 20° | 70 | 63.0 ± 0.7 | 65.5 ± 2.8 | 67.5 ± 3.0 |
| Photonis 25 µm; 60:1; 8° | 60 | 43.5 ± 0.6 | 53.9 ± 4.3 | 55.6 ± 4.6 |

**Table 1** Open area ratio (OAR) and absolute detection efficiency for proton impact for the three investigated MCPs (pore diameter, thickness/pore diameter; pore angle). Hamamatsu ("funnel"): Diameter 80mm, plate thickness 0.72 mm, bias angle 20 ± 1°, center to center spacing max. 15 µm, channel diameter 12 µm, OAR 90 %. Hamamatsu: Diameter 80 mm, plate thickness 0.96mm, bias angle 20 ± 1°, center to center spacing max. 15 µm, channel diameter 12 µm, OAR 70 %. Photonis: Diameter 80 mm, plate thickness 1.5 mm, bias angle 8 ± 1°, center to center spacing max. 32 µm, channel diameter 25 µm, OAR 60 %.

### Direct comparison of "funnel" and traditional MCPs - analyzing the five-particle break-up of CHBrClF

A direct comparison of the efficiency properties of two MCPs requires identical experimental conditions. For this purpose, a symmetric COLTRIMS spectrometer consisting of two identical ion arms (21 cm acceleration length and E = 119 V/cm electric field) was built. On both sides a detector with hexagonal delay-line anode [1] is mounted, one equipped with a Photonis MCP (OAR specified 60 %, slightly used), one with a Hamamatsu MCP (OAR specified 90 %). For further amplification, a second MCP (Photonis) was mounted behind the first. Just as in the experiment presented in the previous section, the ions gained a kinetic energy of approx. 2.5 keV due to the length of the spectrometer and its comparably high electric field. Therefore, no meshes needed to be installed in front of the MCP, which is typically done for post-acceleration of the ions in order to increase the MCP quantum efficiency. The main chamber was baked for one week at 90°C, resulting in a residual gas pressure without gas jet of $1 \cdot 10^{-8}$ Pa. In order to characterize both MCPs, the experiment was performed twice by switching the direction of the electric field of the spectrometer and thus using either one or the other detector. The ionization of the CHBrClF target molecules was induced by focussing short, intense, linearly polarized laser pulses (f = 60 mm, 40 fs, central wave length 800 nm, 1.1 W), generated by a Ti:Sapphire regenerative amplifier (KMLabs Wyvern 500), resulting in a focal intensity of $1.1 \cdot 10^{15}$ W/cm² onto the supersonic gas jet. The jet was produced by expanding CHBrClF with its vapor pressure at room temperature (approx. $6 \cdot 10^4$ Pa) through a nozzle of 30 µm diameter into vacuum.

As test reaction, CHBrClF molecules were multiply ionized by an intense short laser pulse. This leads to a removal of up to five electrons from the molecular system. Subsequently the multiply ionized molecules underwent Coulomb explosion and we detected the fragment ions.

For the purpose of examining the single-particle efficiency, the singly ionized parent ion was analyzed by gating on suitable ranges of the time-of-flight spectrum depicted in Fig. 6 a). Cl and Br, commonly exist as two isotopes, $^{35}$Cl and $^{37}$Cl (76 % and 24 %), respectively $^{79}$Br and $^{81}$Br, (51 % and 49 %). The TOF distribution consists of three peaks which stem from the three possible isotopic mass combinations of the molecular ion (CH$^{79}$Br$^{35}$ClF, CH$^{79}$Br$^{37}$ClF or CH$^{81}$Br$^{35}$ClF, CH$^{81}$Br$^{37}$ClF). The various break-up channels occurring after Coulomb explosion were analyzed by gating on TOF-coincidence maps as shown in Figs. 6 b,c). In Fig. 6 b) the break-up channel into CHClF$^+$ + Br$^+$ is depicted. The time-of-flight of the first ionic fragment is plotted versus that of the second. In this representation (known as Photo Ion / Photo Ion Coincidence (PIPICO or PI2CO) plot, breakup channels of different mass over charge ratio occur as distinct lines. For breakup channels leading to more than two fragments, similar spectra can be obtained by plotting time-of-flight sums against each other. In Fig. 6 c) a corresponding coincidence spectrum for the five body fragmentation is exemplarily presented. There even a tiny contribution of molecules consisting of a $^{13}$C atom is visible. These contributions are labelled as "1". The other labels in Fig. 6 c) assign the measured lines to their corresponding molecular isotopes.

For further analysis the following ionization/breakup channels were selected (integrated over all isotopes):

TOF:     CHBrClF → CHBrClF$^+$
PI2CO:   CHBrClF → CHClF$^+$ + Br$^+$
PI3CO:   CHBrClF → CHF$^+$ + Br$^+$ + Cl$^+$
PI4CO:   CHBrClF → CH$^+$ + Br$^+$ + Cl$^+$ + F$^+$
PI5CO:   CHBrClF → C$^+$ + H$^+$ + Br$^+$ + Cl$^+$ + F$^+$

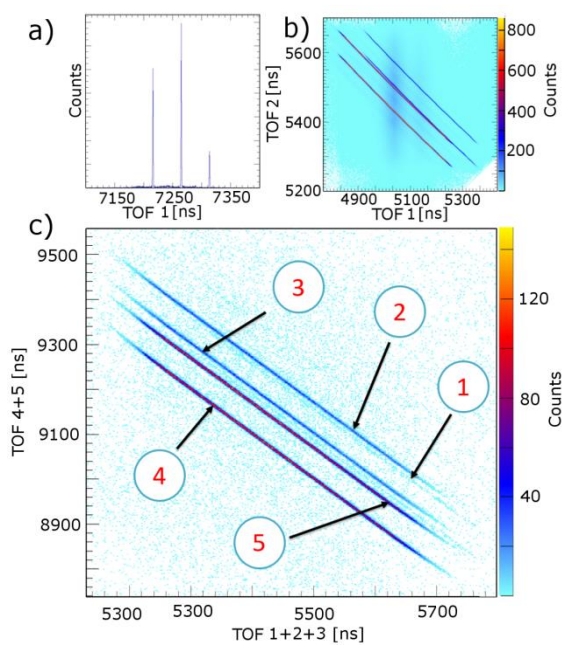

**Fig. 6** a) Time-of-flight spectrum of the parent ion. The different lines correspond to the different combinations of isotopes of Cl and Br. b) Time of Flight coincidence map for the fragmentation of CHBrClF into CHFCl$^+$ and Br$^+$. c) TOF-coincidence map for the fragmentation of CHBrClF into five particles, measured with the "funnel" MCP. The different lines are isotopic lines. The horizontal axis shows the sum of the TOFs of the particles 1 to 3 (numbered by the time order in which they hit the detector). The vertical axis shows the sum of the TOFs of particle 4 and 5. Line number 1 highlights a break-up channel with $^{13}$C ions ($^{13}$C$^+$+ H$^+$+ $^{79}$Br$^+$+ $^{35}$Cl$^+$+ F$^+$). Line number 2 corresponds to the isotopic break-up into $^{12}$C$^+$+ H$^+$+ $^{81}$Br$^+$+ $^{37}$Cl$^+$+ F$^+$, line number 3 into $^{12}$C$^+$+ H$^+$+ $^{79}$Br$^+$+ $^{37}$Cl$^+$+ F$^+$, line number 4 into $^{12}$C$^+$+ H$^+$+ $^{79}$Br$^+$+ $^{35}$Cl$^+$+ F$^+$, and line number 5 into $^{12}$C$^+$+ H$^+$+ $^{81}$Br$^+$+ $^{35}$Cl$^+$+ F$^+$. The isotopes of hydrogen and fluorine ions correspond to $^1$H and $^{19}$F, respectively. The total measurement time for the "funnel" MCP was about 27 min.

Table 2 summarizes the number of detected events obtained for the different ionization channels for each of the two MCPs under test. The values listed for the "standard" MCP have been normalized by comparing the single ionization case of both MCPs.

The probability to detect all fragments of a break-up channel is composed of the probability that this break-up occurs (p$_{break-up}$) and the single-particle detection efficiency ε of the detector to the power of the number of detected particles $n$. By normalizing the events measured with the "standard" MCP to same single ion count rate of the "funnel" MCP, the longer measuring time for the "standard" MCP is compensated.

We have to note that a direct comparison of the results to those presented in the previous section is unfortunately not possible since the quantum efficiency of the MCPs depends – (assuming same ion impact energy)- in particular on the ion mass **[4]**. Additionally, different MCPs can have a differing mass to efficiency dependency.

|  | Photonis ("standard") 25 µm; 60:1; 8° | Hamamatsu („funnel") 12 µm; 80:1; 20° |
|---|---|---|
| TOF | 463007 | 463007 |
| PI2CO | 184682 | 263722 |
| PI3CO | 19960 | 57917 |
| PI4CO | 1330 | 8736 |
| PI5CO | 122 | 1677 |

**Table 2** Yield for the "standard" and "funnel" MCP for different break-up channels of CHBrClF, the numbers for the standard MCP were normalized to the same single ion count rate of the "funnel" MCP.

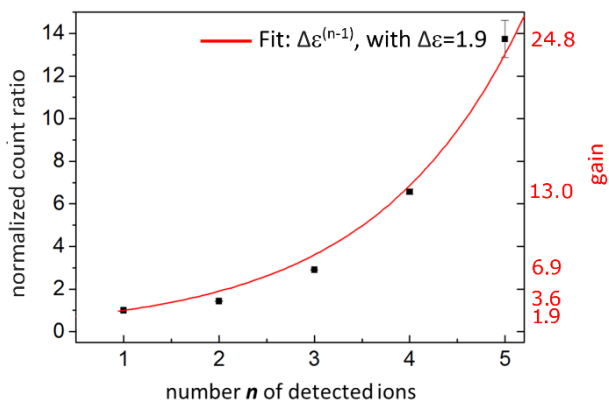

**Fig. 7** Count ratio plotted versus the number $n$ of ions detected in coincidence. An exponential fit reveals the dramatic efficiency enhancement between the Photonis "standard" (OAR 60 %) and the Hamamatsu "funnel" (OAR 90%) MCP. For n = 1-4 the error bars are within the dot size. The "gain" is calculated by $1.9^n$.

## Conclusion

In this article, three different types of MCPs have been investigated in order to directly determine their absolute detection efficiency: Hamamatsu's "funnel" MCP with an OAR of 90 % (ε = 86.0 ± 1.2), the "standard" MCP by Hamamatsu with an OAR of 70 % (ε = 67.5 ± 3.0) and the "standard" MCP by Photonis with an OAR of 60 % (ε = 55.6 ± 4.6). These ideal efficiency values, however, cannot be achieved under typical experimental conditions as additional experimental parameters deteriorate the efficiency. The most important ones are electronic noise and (in particular for heavier ions) ion impact energies that are insufficient to reach saturation.

In a second experiment, we demonstrated the importance of the detector efficiency when performing (multi-)coincidence measurements. Normalized to the same single particle detection rate, the "funnel" MCP recorded approx. 24 times more five-particle break-ups than a "normal" MCP. An increase in yield of such magnitude corresponds, for example, to a reduction of measuring time from one day to one hour.

Two further observations concerning "funnel" MCPs in coincidence set-ups such as COLTRIMS reaction microscopes were made. Firstly, we observed an increase of the residual gas pressure during operation of the "funnel" MCPs probably due to outgassing inside the pores induced by the electron avalanche. We cannot precisely specify the outgas rate, but its magnitude was in the range of $10^{-4}$ Pa•l/s. The outgassing decreased during operation. Secondly, we observed no excessive wear of the MCP over four weeks of continuous use at count rates of 20-30 kHz (rather homogenously distributed across the MCP).

## Acknowledgement

We acknowledge support from Deutsche Forschungsgemeinschaft via Sonderforschungsbereich 1319 (ELCH). acknowledges support by the German National Merit Foundation. M. S. thanks the Adolf-Messer foundation for financial support.

## References

[1] O. Jagutzki et al., "Multiple hit readout of a microchannel plate detector with a three-layer delay-line anode," *IEEE Trans. Nucl. Sci.*, vol. 49, no. 5, pp. 2477–2483, 2002.

[2] J. H. D. Eland et al., "Complete two-electron spectra in double photoionization: The rare gases Ar, Kr, and Xe," (eng), *Physical review letters*, vol. 90, no. 5, p. 53003, 2003.

[3] R. Dörner et al., "Cold Target Recoil Ion Momentum Spectroscopy: A 'momentum microscope' to view atomic collision dynamics," *Physics Reports*, vol. 330, no. 2, pp. 95–192, http://www.sciencedirect.com/science/article/pii/S037015739900109X, 2000.

[4] M. Krems, J. Zirbel, M. Thomason, and R. D. DuBois, "Channel electron multiplier and channelplate efficiencies for detecting positive ions," *Review of Scientific Instruments*, vol. 76, no. 9, p. 93305, 2005.

[5] B. Brehm, J. Grosser, T. Ruscheinski, and M. Zimmer, "Absolute detection efficiencies of a microchannel plate detector for ions," *Measurement Science and Technology*, vol. 6, no. 7, p. 953, http://stacks.iop.org/0957-0233/6/i=7/a=015, 1995.

[6] A. Vredenborg, W. G. Roeterdink, and M. H. M. Janssen, "A photoelectron-photoion coincidence imaging apparatus for femtosecond time-resolved molecular dynamics with electron time-of-flight resolution of sigma=18 ps and energy resolution Delta E/E=3.5%," (eng), *Review of Scientific Instruments*, vol. 79, no. 6, p. 63108, 2008.

[7] S. Matoba, R. Takahashi, C. Io, T. Koizumi, and H. Shiromaru, "Absolute Detection Efficiency of a High-Sensitivity Microchannel Plate with Tapered Pores," *Jpn. J. Appl. Phys.*, vol. 50, p. 112201, 2011.

[8] S. Matoba et al., "Note: Absolute detection efficiency of a tapered microchannel plate for Ne$^+$ ions," (eng), *The Review of scientific instruments*, vol. 85, no. 8, p. 86105, 2014.

[9] A. L. Bennani, J. Pebay, and B. Nguyen, "Measurement of the absolute electron detection efficiency of a channel multiplier (channeltron)," *Journal of Physics E: Scientific Instruments*, vol. 6, no. 11, p. 1077, http://stacks.iop.org/0022-3735/6/i=11/a=004, 1973.

[10] F. Bordoni, "Channel electron multiplier efficiency for 10–1000 eV electrons," *Nuclear Instruments and Methods*, vol. 97, no. 2, pp. 405–408, http://www.sciencedirect.com/science/article/pii/0029554X71903004, 1971.

[11] F. Bordoni, M. Martinelli, M. P. Fioratti, and S. R. Piermattei, "Absolute efficiency of channeltron electron multipliers for 10–100 keV X-rays," *Nuclear Instruments and Methods*, vol. 116, no. 1, pp. 193–194, http://www.sciencedirect.com/science/article/pii/0029554X74906004, 1974.

[12] S. Hosokawa, N. Takahashi, M. Saito, and Y. Haruyama, "Absolute detection efficiencies of a

microchannel plate detector for 0.5-5 keV neutrals," (eng), *The Review of scientific instruments*, vol. 81, no. 6, p. 63301, 2010.

[13] A. Müller, N. Djurić, G. H. Dunn, and D. S. Belić, "Absolute detection efficiencies of microchannel plates for 0.1–2.3 keV electrons and 2.1–4.4 keV Mg + ions," *Review of Scientific Instruments*, vol. 57, no. 3, pp. 349–353, 1986.

[14] H. C. Straub, M. A. Mangan, B. G. Lindsay, K. A. Smith, and R. F. Stebbings, "Absolute detection efficiency of a microchannel plate detector for kilo-electron volt energy ions," *Review of Scientific Instruments*, vol. 70, no. 11, pp. 4238–4240, 1999.

[15] B. Gaire *et al.,* "Determining the absolute efficiency of a delay line microchannel-plate detector using molecular dissociation," *Review of Scientific Instruments*, vol. 78, no. 2, p. 24503, 2007.

[16] N. Takahashi, S. Hosokawa, M. Saito, and Y. Haruyama, "Corrigendum: Measurement of absolute detection efficiencies of a microchannel plate using the charge transfer reaction," *Physica Scripta*, vol. 86, no. 4, p. 49501, http://stacks.iop.org/1402-4896/86/i=4/a=049501, 2012.